\documentclass[journal]{IEEEtran}
\usepackage{cite}
\usepackage{amsmath,amssymb,amsfonts}
\usepackage{algorithmic}
\usepackage{graphicx}
\usepackage{textcomp}
\usepackage{wrapfig}
\def\BibTeX{{\rm B\kern-.05em{\sc i\kern-.025em b}\kern-.08em
    T\kern-.1667em\lower.7ex\hbox{E}\kern-.125emX}}
\usepackage{bm}
\newcommand{\ee}        {{\rm e}}
\newcommand{\jj}        {{\rm j}}
\newcommand{\dd}        {{\rm d}}
\newcommand{\TT}        {{\rm T}}
\newcommand{\HH}        {{\rm H}}

\begin{document}
\title{Radar Positioning for Accurate Sensing of Pulse Waves at Multiple Sites Using a 3D Human Model}
\author{Takehito Koshisaka and Takuya Sakamoto, \IEEEmembership{Senior Member, IEEE}
\thanks{Manuscript received April 16, 2023;
revised May 31, 2024; accepted May 31, 2024. Date of publication May 31, 2024; date of current version May 31, 2024.} 
\thanks{This work was supported in part by JST under Grants JPMJCE1307 and JPMJMI22J2, by SECOM Science and Technology Foundation, and JSPS KAKENHI under Grants 19H02155, 21H03427, and 23H01420.}
\thanks{T.~Koshisaka and T.~Sakamoto are with the Department of Electrical Engineering, Graduate School of Engineering, Kyoto University, Kyoto, Kyoto 615-8510, Japan (e-mail: sakamoto.takuya.8n@kyoto-u.ac.jp).}}

\maketitle

\begin{abstract}
  This study proposes a sensing method using a millimeter-wave array radar and a depth camera to measure pulse waves at multiple sites on the human body. Using a three-dimensional shape model of the target human body measured by the depth camera, the method identifies reflection sites on the body through electromagnetic scattering simulation. On the basis of the simulation, the radar system can be positioned at a suitable location for measuring pulse waves depending on the posture of the target person. Through measurements using radar and depth camera systems, we demonstrate that the proposed method can estimate the body displacement waveform caused by pulse waves accurately, improving the accuracy by 14\% compared with a conventional approach without a depth camera. The proposed method can be a key to realizing an accurate and noncontact sensor for monitoring blood pressure.
\end{abstract}

\begin{IEEEkeywords}
Electromagnetic scattering, pulse wave, millimeter-wave radar, depth camera
\end{IEEEkeywords}


\section{Introduction}
\label{sec:introduction}
\IEEEPARstart{H}{ypertension} increases the risk of cardiovascular diseases \cite{Vasan2001}, and the percentage of hypertensive patients aged 30--79 years was 33.1\% in 2019 \cite{WHO}. Because regular blood pressure measurement is necessary to keep blood pressure within an appropriate range \cite{Bryant2020}, it is necessary to have a sensing system for constant monitoring of blood pressure.

Blood pressure is related to pulse transit time, which is calculated from the arrival time of pulse waves measured at multiple sites on the human body \cite{Mukkamala2015}. This has been achieved mostly using contact sensors \cite{Chao2021,He2014,Roy2022}. Research on noncontact pulse wave measurement has been conducted intensively to make the system unobtrusive and easy to set up; the use of millimeter-wave radar systems has been reported for such purposes \cite{Michler2019,Michler2020,Buxi2017,Ebrahim2019,Zhao2018,Johnson2019,Lu2010,Oyamada2021}. For example, Oyamada \emph{et al.} measured the body displacement waveform at the back and thigh simultaneously using a 79-GHz frequency-modulated continuous-wave (FMCW) radar system, and the accuracy was evaluated through simultaneous measurements with laser displacement sensors. To apply this technique in practice, we need to adjust the position of the radar system or the target person to receive echoes from the target sites on the body.

In this study, we propose a new approach for determining the position of the radar system depending on the target person's position and posture using electromagnetic scattering simulation to estimate which part of the body the reflected waves are received from. We reduce the computational complexity using a simplified approximation instead of full-wave electromagnetic simulation. Some studies on electromagnetic simulations with a human body are also based on the physical optics (PO) approximation because of its simplicity and practicality  \cite{Ranjbar2014,Ranjbar2015,Mazzinghi2018,Zhang2021,Arias2019,Corucci2012}. Some studies use geometrically simplified human models \cite{Manfredi2019,Keerativoranan2020}, while other studies use precise human models obtained from a 3D laser scanner \cite{Perez-Eijo2019,Mokhtari-Koushyar2014,Vahidpour2012}.

Konishi \emph{et al.} reported electromagnetic wave scattering analysis with a human model obtained by a depth camera, assuming a radar frequency of 60 GHz, and visualized the reflection sites on the human body \cite{Konishi2018,Sakamoto-Konishi2018}. Although the method proposed by Konishi \emph{et al.} enables estimation of reflection sites on the human body surface, the method has not been applied to pulse wave measurement. In this study, we propose a method to achieve accurate pulse wave measurement without prior adjustment of the participant's body posture by estimating the reflection sites on the target human body using electromagnetic wave scattering analysis to determine an appropriate radar position. 

To obtain 3D human body shape data, Vahidpour and Sarabandi \cite{Vahidpour2012} used a 3D laser scanner to obtain a precise human model, which requires several cameras and a sufficiently large space. In this study, we use a depth camera to acquire a human model following the approach of Konishi \emph{et al.} \cite{Konishi2018,Sakamoto-Konishi2018}. Although some studies have used both a depth camera and a radar system \cite{Shokouhmand2022,Mikhelson2012}, they used the depth data only to estimate the position of the target person, whereas we consider the shape of the human body for radar measurement. 

In this study, we propose a system to measure pulse waves at multiple locations with high accuracy by combining a depth camera and millimeter-wave radar. In Section \ref{sec:2}, we investigate the characteristics of millimeter-wave scattering from the human body surface by performing electromagnetic wave scattering analysis based on the PO approximation using a precise human model. In Section \ref{sec:3}, we propose a measurement system that integrates a depth camera and a millimeter-wave radar system, and evaluate its performance through experiments with two participants in the supine and prone positions.

\section{Electromagnetic Scattering Analysis and Estimation of Reflection Sites}
\label{sec:2}
\subsection{Electromagnetic Scattering Analysis with the PO Approximation}
The PO approximation is used for electromagnetic wave scattering analysis. This approximation calculates the current distribution on the target surface caused by the incident magnetic field and the normal vector of the surface. If the target material is modeled as a perfect electric conductor, the current density $\bm{J}(\bm{r}')$ at position $\bm{r}'$ on the target surface is approximated as
\begin{equation}
    \bm{J}(\bm{r}') = 2\hat{\bm{n}}(\bm{r}') \times \bm{H}^\mathrm{inc}(\bm{r}'), \label{eq:1}
\end{equation}
where $\hat{\bm{n}}(\bm{r}')$ is a unit normal vector, and $\bm{H}^\mathrm{inc}(\bm{r}')$ is an incident magnetic field. 

Assuming that the polarization directions of the transmitting and receiving elements are both in the $z$-direction and the current density $\bm{J}(\bm{r}')$ at each position $\bm{r}'$ is approximated by an infinitesimal dipole, the $z$-component of the radiated electric field ${E}_z(\bm{r};\bm{r}')$ at antenna position $\bm{r}$ is approximated as
\begin{equation}
    {E}_z(\bm{r};\bm{r}') = \jj\dfrac{k Z_0\ell}{4\pi|\bm{r}-\bm{r}'|}       {\hat{\bm{z}}\cdot\bm{J}(\bm{r}')}\ee^{-\jj k|\bm{r}-\bm{r}'|},  \label{eq:2}
\end{equation}
where $\hat{\bm{z}}$ is a unit $z$ vector, $\jj$ is an imaginary unit, $\ell$ is the length of the dipole, $Z_0$ is the impedance of free space, and $k$ is the wave number. Here we assume $|\bm{r}-\bm{r}'|$ is sufficiently large compared with the reflection area.

Shijo \emph{et al.}\cite{Shijo2004} proposed a PO-based method for visualizing high-frequency diffraction using a weighting function, called the eye function, defined as
\begin{equation}
    w(\bm{r};\bm{r}') = 
    \begin{cases}
        \dfrac{1}{2} \left\{\cos{\left(\dfrac{\pi|\bm{r}-\bm{r}'|}{a_0}\right)} + 1\right\} & (|\bm{r}-\bm{r}'|\leq a_0) \\
        0 & (\mathrm{otherwise})
    \end{cases},\label{eq:3}
\end{equation}
where $a_0$ is the radius of the eye function and should be sufficiently larger than the wavelength \cite{Shijo2004}. Using \eqref{eq:3}, we can approximate the intensity of the electric field received from the vicinity of the scattering point $\bm{r}'$ as
\begin{equation}
    E_\mathrm{scat}(\bm{r}') = \left|\iint_{\Omega} w(\bm{r};\bm{r}') E_z(\bm{r};\bm{r}')\,\dd S\right|^2, \label{eq:4}
\end{equation}
where $\Omega$ is the entire target boundary surface.

\begin{figure}[!t]
\centerline{\includegraphics[width=0.9\columnwidth]{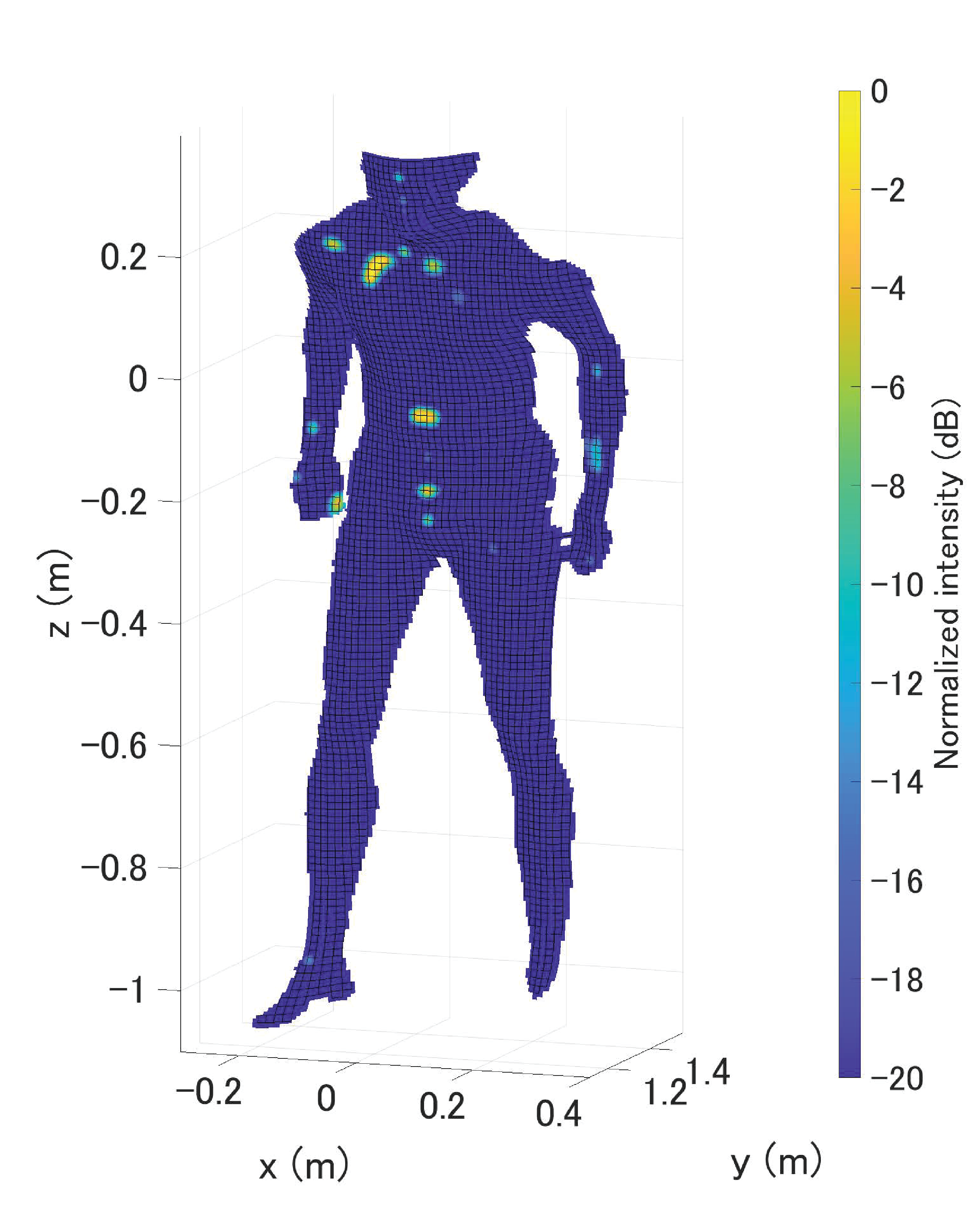}}
\caption{Intensity of the reflected electric field distribution for a human model in a supine position calculated with PO for 79~GHz.}
\label{fig:1}
\end{figure}

A human 3D model is obtained using an iReal 2E laser scanner (SCANTECH Co., Ltd., Hangzhou, China), whose parameters are given in Table \ref{table:1}. Fig. \ref{fig:1} shows $E_\mathrm{scat}(\bm{r}')$ for the human model in a supine position, where we set $a_0=5\lambda$ and $\lambda$ is the wavelength. In the figure, we see that reflection occurs mainly at several spots around the chest and abdomen. In addition, each area contributing reflection is small compared to the size of the human body.

\begin{table}[tb]
    \centering
    \caption{Parameters of the laser scanner}
    \label{table:1}
    \begin{tabular}{|c|c|} \hline
    Light source & Infrared \\ \hline
    Repeatability & 0.1 mm \\ \hline
    Resolution & 0.2 to 3 mm \\ \hline
    \end{tabular}
\end{table}

\subsection{Simplified Method for Estimation of Reflection Spots}
\label{subsec:2B}
Because the calculation of $E_\mathrm{scat}$ using the PO approximation according to Eq.~(\ref{eq:4}) contains a double integral over an area much larger than the wavelength, the computational cost can be unacceptably large in practice. To address this issue, we introduce an even simpler method as follows: Let $\bm{r}$ be a radar antenna position and $\bm{r}'$ be a position on the target surface. We define
\begin{equation}
    \xi(\bm{r};\bm{r}') = \cos^{-1}\left( \dfrac{\hat{\bm{n}}(\bm{r}')\cdot(\bm{r}-\bm{r}')}{|\bm{r}-\bm{r}'|} \right), \label{eq:5}
\end{equation}
where $\xi$ is the angle between the radar line-of-sight direction 
and the normal vector $\hat{\bm{n}}$. Note that $\xi$ satisfies $0\leq\xi\leq\pi$.

\begin{figure}[!t]
\centerline{\includegraphics[width=0.9\columnwidth]{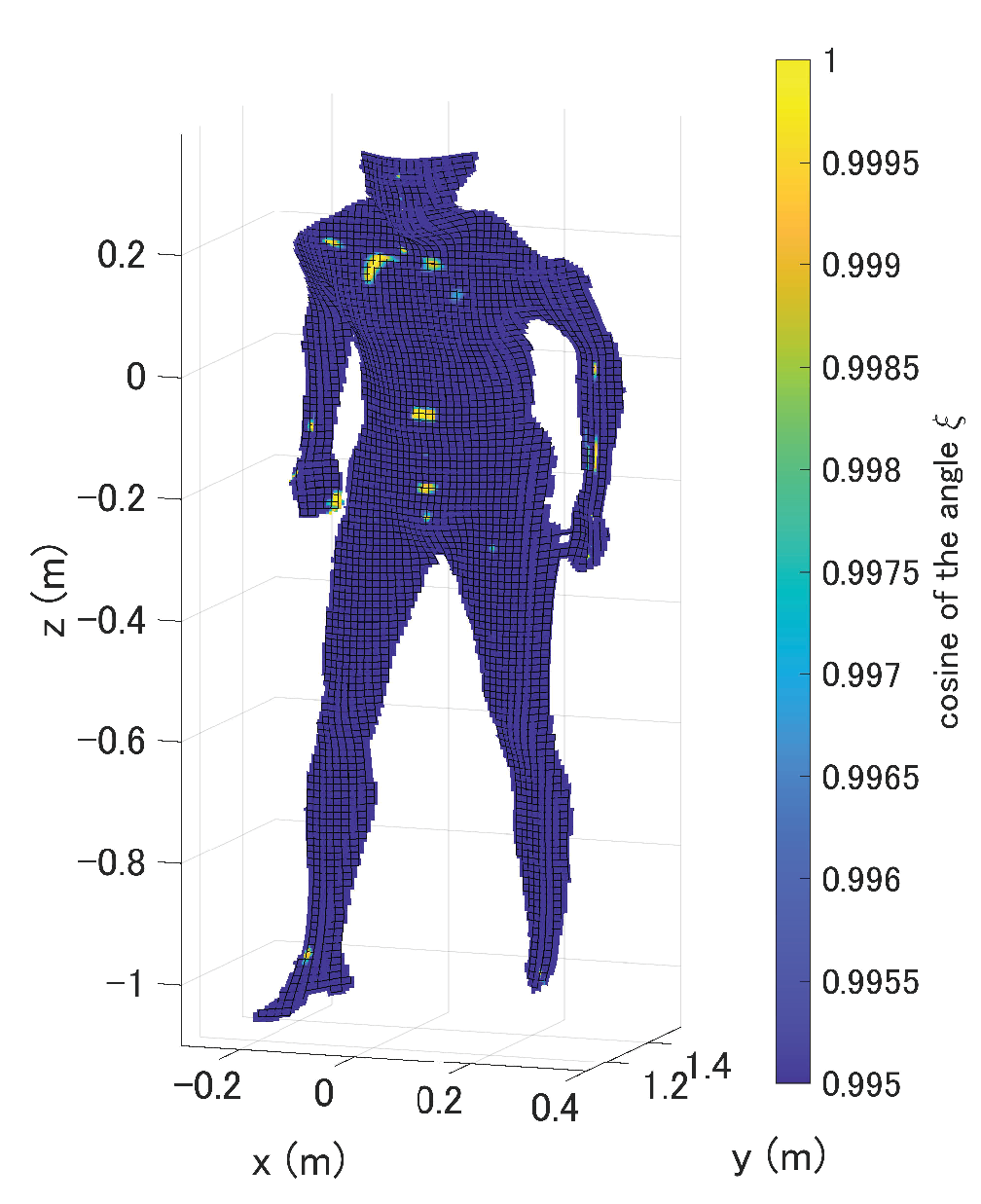}}
\caption{Cosine of the angle between the radar line-of-sight vector and the normal vector.}
\label{fig:2}
\end{figure}

If a point located at $\bm{r}'$ on the target surface has a large $|\xi|$, the electric field scattered from the point is unlikely to contribute to the radar signal. Fig. \ref{fig:2} shows $\cos\xi$ for a human model in a supine position with the same condition as in Fig.~\ref{fig:1}. Comparing this with Fig. \ref{fig:1}, we see that $\cos{\xi}$ is large close to the reflection area on the target surface, which indicates that $\cos\xi$ can be used approximately to estimate the reflection areas instead of the PO approximation, and the computational cost in calculating $\xi$ is much smaller than that of PO-based calculation.

\section{Proposed Measurement System and Performance Evaluation}
\label{sec:3}
\subsection{Experimental Setup}
In this study, we propose a method to determine an appropriate location of the radar system to obtain echoes from target sites on the human body. The method assumes a system with a millimeter-wave radar and depth camera installed over a target person lying in a prone position on a bed.
We use an Intel RealSense D435 depth camera (Intel Corp., Santa Clara, California, U.S.) instead of the laser scanner used in the previous section. The parameters of the depth camera are given in Table \ref{tab:2}. Although the precision of a human model obtained with a depth camera is lower than that obtained with a laser scanner, a depth camera can acquire a human model in a shorter time and is suitable in practice.

\begin{table}[b]
    \centering
    \caption{Parameters of the depth camera}
    \label{tab:2}
    \begin{tabular}{|c|c|} \hline
        Resolution & $1280\times 720$ \\ \hline
        Frame rate & $30\,\mathrm{fps}$ \\ \hline
        Horizontal field of view & $87^\circ \pm 3^\circ$ \\ \hline
        Vertical field of view & $58^\circ \pm 1^\circ$ \\ \hline
        Diagonal field of view & $95^\circ \pm 1^\circ$ \\ \hline
    \end{tabular}
\end{table}

\begin{figure}[!t]
\centerline{\includegraphics[width=0.9\columnwidth]{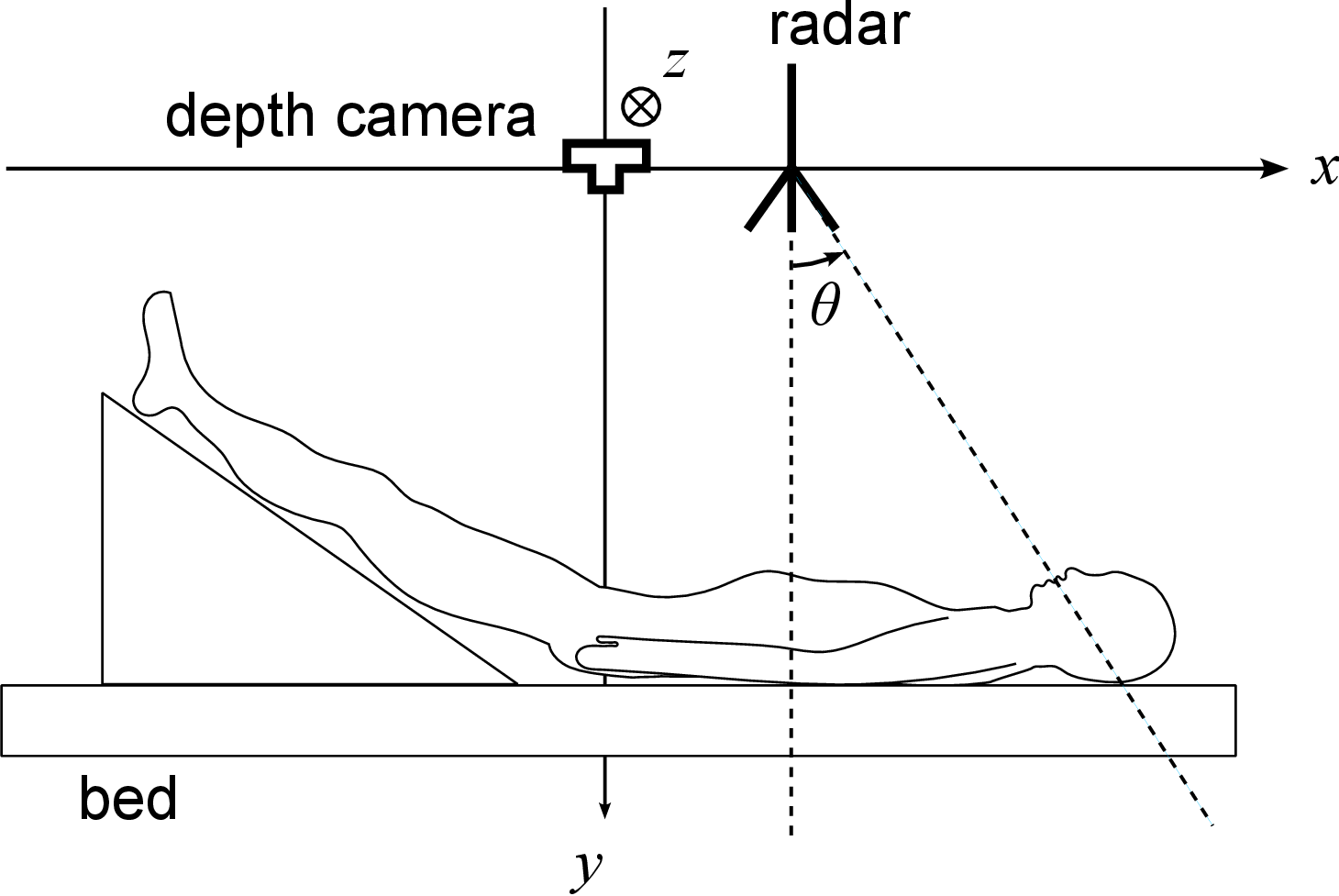}}
\caption{Schematics of the proposed measurement system with a depth camera located at the origin and a radar system on the $x$-axis.}
\label{fig:3}
\end{figure}

Fig. \ref{fig:3} shows a diagram of the proposed measurement system with a coordinate system with the $x$-axis in the direction of the body axis and the $y$-axis vertical to the floor. The depth camera is fixed at the top of the bed and set as the origin of the coordinate system. The millimeter-wave radar can scan along the $x$-axis, where the $x$-axis is 1.1~m from the bed top.

The measurement procedure using the proposed method is as follows:
\begin{enumerate}
    \item Obtain a human 3D model using a depth camera.
    \item Visualize the radiated power distribution.
    \item Locate the radar position to obtain echoes from the desired sites.
    \item Measure a body displacement waveform using the radar system.
\end{enumerate}

In this study, we used a radar module (T14\_01120112\_2D, S-Takaya Electronics Industry, Okayama, Japan) that comprises a single-chip millimeter-wave sensor (IWR 1443, Texas Instruments Inc., Dallas, Texas, United States). The parameters of the radar module are given in Table \ref{tab:3}. We used a multiple-input and multiple-output (MIMO) array radar. If the distance to a target from the radar antennas is sufficiently greater than the antenna aperture size, the array can be approximated as a virtual array with $M=M_\mathrm{T}M_\mathrm{R}=12$ elements, where $M_\mathrm{T}=3$ and $M_\mathrm{R}=4$ are the numbers of transmitting and receiving elements, respectively. The element spacing of the virtual array of the radar module was half the wavelength, and the angular resolution was approximately $8.5^\circ$. We used laser displacement sensors (CDX-150, OPTEX FA Co., Ltd., Kyoto, Japan) as a reference for body displacement measurement. The parameters of the laser displacement sensors are shown in Table \ref{tab:4}.

\begin{table}[tb]
    \centering
    \caption{Parameters of the millimeter-wave radar}
    \label{tab:3}
    \begin{tabular}{|c|c|} \hline
    Type of radar & FMCW radar \\ \hline
    Center frequency & 79.0~GHz \\ \hline
    Bandwidth & 3.4~GHz \\ \hline
    Number of transmit antennas & 3 \\ \hline
    Number of receive antennas & 4 \\ \hline
    Spacing of transmit antennas & 7.6~mm ($2\lambda$) \\ \hline
    Spacing of receive antennas & 1.9~mm ($\lambda/2$) \\ \hline
    Beamwidth in the E-planes & $\pm 4^\circ$   \\ \hline
    Beamwidth in the H-planes & $\pm 35^\circ$ \\ \hline
    Range resolution & 44~mm \\ \hline
    Sampling frequency (slow-time) & 145.6~Hz \\ \hline
    \end{tabular}
\end{table}

\begin{table}[tb]
    \centering
    \caption{Parameters of the laser sensor}
    \label{tab:4}
    \begin{tabular}{|c|c|} \hline
    Measurement principle & triangulation \\ \hline
    Wavelength of the light source & 655~nm \\ \hline
    Diameter of the spot size & 120~\textmu m \\ \hline
    Repeatability & 0.2~\textmu m \\ \hline
    Sampling frequency & 200~Hz \\ \hline
    \end{tabular}
\end{table}

\begin{figure}[!t]
\centerline{\includegraphics[width=0.9\columnwidth]{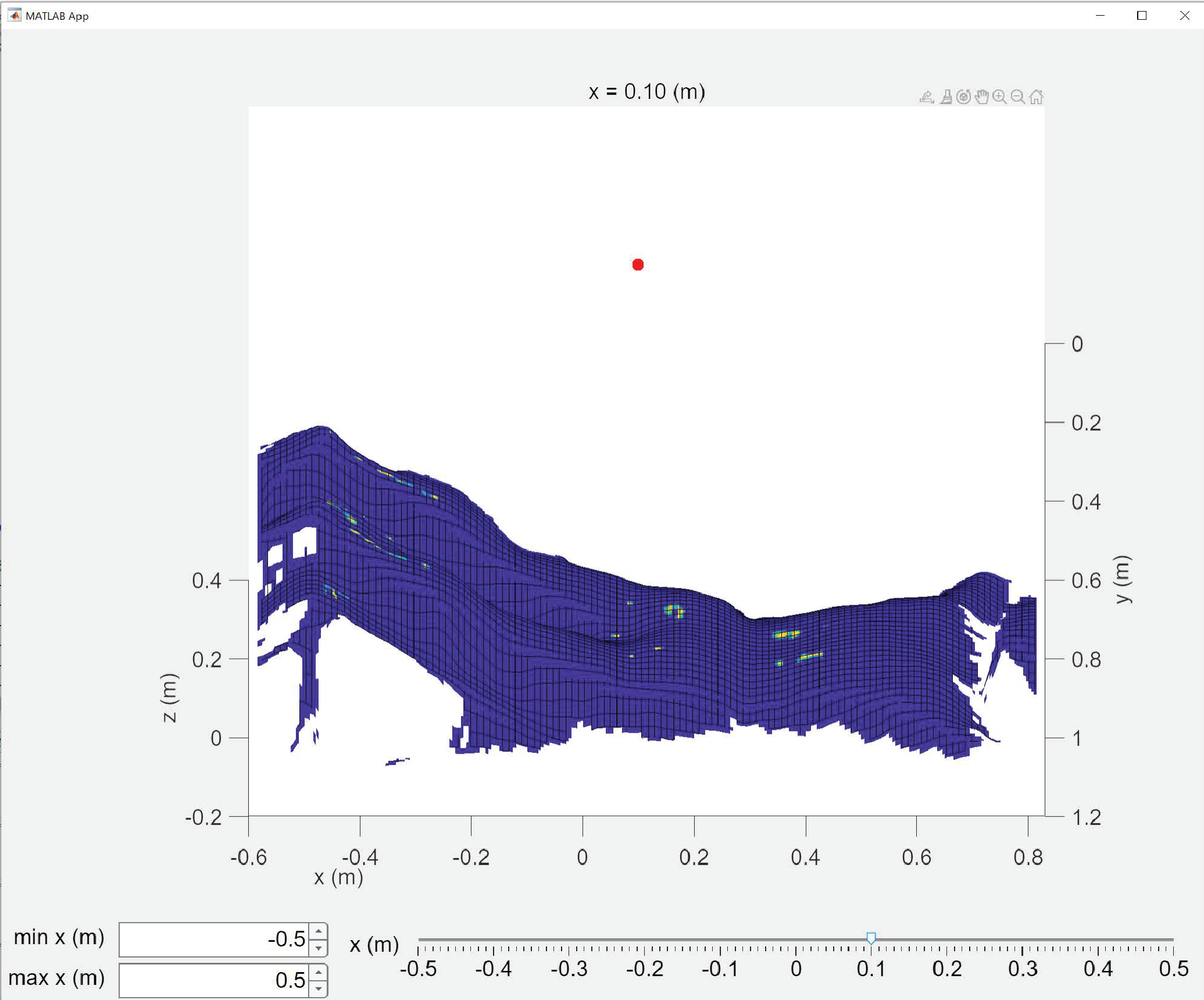}}
\caption{Software for visualizing $\cos\xi$ that indicates the intensity of the reflected electric field, where the red dot represents the position of a radar. Its $x$-coordinate can be changed with the slider.}
\label{fig:4}
\end{figure}

In this study, we developed software that displays an image of the human body with $\cos\xi$ and indicates the intensity of the reflected electric field. With this software, the reflection sites on the body can be confirmed immediately when the radar position changes. The software runs on MATLAB as shown in Fig. \ref{fig:4}, in which the image corresponds to $\cos\xi(\bm{r})$ calculated using the simplified method in Section \ref{subsec:2B}. The $x$-coordinate of the radar position can be changed with the slider at the bottom of the screen. The radar position is expected to be optimized automatically if an appropriate objective function is defined. Despite this, the implementation of the optimization process is out of the scope of this study and will be an important topic to study in the future. In this study, we selected the radar position manually using this software.

\subsection{Array Radar Measurement of Pulse Waves}
We propose a method for measuring a displacement waveform at multiple sites on the human body using a millimeter-wave radar system. 
Let us assume $N$ reflections points on the target human body, and let $\theta_n$ be the direction of arrival of the $n$th echo with the corresponding displacement $d_n(t)$ for $n=1,2,\cdots,N$.
The $n$th echo is phase-modulated by the body motion as 
$s_n(t)\propto \ee^{\jj 2kd_n(t)}$. For simplicity, we assume 
$s_n(t)= \ee^{\jj 2kd_n(t)}$ hereafter. 

For a radar system with a virtual array comprising $M$ elements with an element spacing of $\lambda/2$, the signal received from the $m$th element $x_m(r,t)$ $(m=1,2,\cdots,M)$ is expressed as a function of range $r=c\tau/2$ and slow time $t$, where $\tau$ is a fast time. The received signal $x_m$ is then modeled as
\begin{equation}
    x_m(r,t)=\sum_{n=1}^N a_{m,n}(r) s_n(t),
\end{equation}
where $a_{m,n}(r)$ is a complex constant corresponding to the propagation path between the $n$th target and $m$th antenna element. Then, we define signal vector $\bm{x}(r,t) = [x_1(r,t),...,x_M(r,t)]^\TT$, which is used to estimate the $n$th echo $x_n(r,t)$ as $\hat{x}_n(r,t) = \bm{w}_n^\HH \bm{x}(r,t)$ 
using an appropriate weight vector $\bm{w}_n = [w_{1,n},...,w_{M,n}]^\TT$ for the $n$th echo, where superscripted $\TT$ and $\HH$ represent a transpose operator and a complex-conjugate transpose operator, respectively. By selecting range $r_n$ for the $n$th echo, we can estimate the displacement $d_n(t)$ as $\hat{d}_n(t) = (1/2k)\angle \hat{x}_n(r_n,t)$.

To determine weight vector $\bm{w}$, we use a minimum variance distortionless response (MVDR) method, which is also called directionally constrained minimization of power (DCMP) \cite{Takao1976}. Let $\bm{a}(\theta)$ be a steering vector for direction of arrival $\theta$ expressed as $\bm{a}(\theta)=[a_1(\theta),\cdots,a_M(\theta)]^\TT$, where $a_m(\theta)=\ee^{\jj m\pi \sin{\theta}}$ for an array with antenna spacing $\lambda/2$.

The MVDR (DCMP) weight vector is determined by
\begin{equation}
    \bm{w}(\theta) = \left(\bm{a}^\HH(\theta) R_{xx}^{-1} \bm{a}(\theta)\right)^{-1} R_{xx}^{-1} \bm{a}(\theta),
\end{equation}
where $R_{xx}$ is a correlation matrix of $\bm{x}$ formulated as $R_{xx}=\mathrm{E}[\bm{x}(r_n,t)\bm{x}^\HH(r_n,t)]\simeq (1/T) \int_0^T \bm{x}(r_n,t)\bm{x}^\HH(r_n,t)~\dd t$ with the measurement time length $T$. 

For example, when measuring the displacements at two sites $(r,\theta)=(r_1,\theta_1)$ and $(r_2,\theta_2)$ on the target body, we used the MVDR weight vectors $\bm{w}(\theta_1)$ and $\bm{w}(\theta_2)$ to estimate the displacements $\hat{d}_1(t) = (1/2k)\angle \{\bm{w}^\HH(\theta_1)\bm{x}(r_1,t)\}$ and $\hat{d}_2(t) = (1/2k)\angle \{\bm{w}^\HH(\theta_2)\bm{x}(r_2,t)\}$.

\subsection{Experimental Evaluation of the Proposed Method}
We performed experiments using a depth camera, a millimeter-wave radar system, and a pair of laser displacement sensors with two participants lying on a bed in supine and prone positions. The measurement was performed six times for each participant. For three out of the six times, the position of the radar system was adjusted using the proposed method (case A), and for the other three times, the position was not adjusted (case B). The radar position in case B was shifted by $0.3$ m from the position in case A. The target sites on the body were selected to be the chest and the front of the thigh when the participants were in supine positions, and were selected to be the back and the back of the thigh when the participants were in prone positions. The radar and laser measurements were performed simultaneously for $30$ s while the participants were holding their breaths to prevent respiratory effects.

The accuracy of the proposed method is evaluated in terms of the correlation coefficient $\rho_n$ of the estimated and reference displacement waveforms $\hat{d}_n(t)$ and $d_n(t)$ defined as
\begin{equation}
    \rho_n = \frac{1}{\hat{D}_n D_n}\int_{0}^{T} \hat{d}_n(t) d_n(t)\,\dd t, \label{eq:12}
\end{equation}
where $\hat{D}_n^2 =\int_0^T \hat{d}_n^2(t)\,\dd t$ and ${D}_n^2 =\int_0^T {d}_n^2(t)\,\dd t$ assuming $\int_0^T\hat{d}(t)\dd t=\int_0^T d(t)\dd t =0$ for $T=5.0$ s.

We also evaluate the root-mean-square (RMS) error $\varepsilon_n$ between $\hat{d}_n(t)$ and $d_n(t)$ defined as
\begin{equation}
    \varepsilon_n^2 = \dfrac{1}{T} \int_{0}^{T} |\hat{d}_n(t) - \alpha d_n(t)|^2\,\dd t  \label{eq:13}
\end{equation}
where $\alpha$ is adjusted properly to compensate for the inclination of the laser displacement sensors. The range $r_n$ for the measurement and definition of slow time $t=0$ are also adjusted to compensate for the imperfect synchronization of the radar and laser data.

\begin{figure}
    \centering
    \begin{minipage}{0.49\columnwidth}
      \includegraphics[width=\columnwidth]{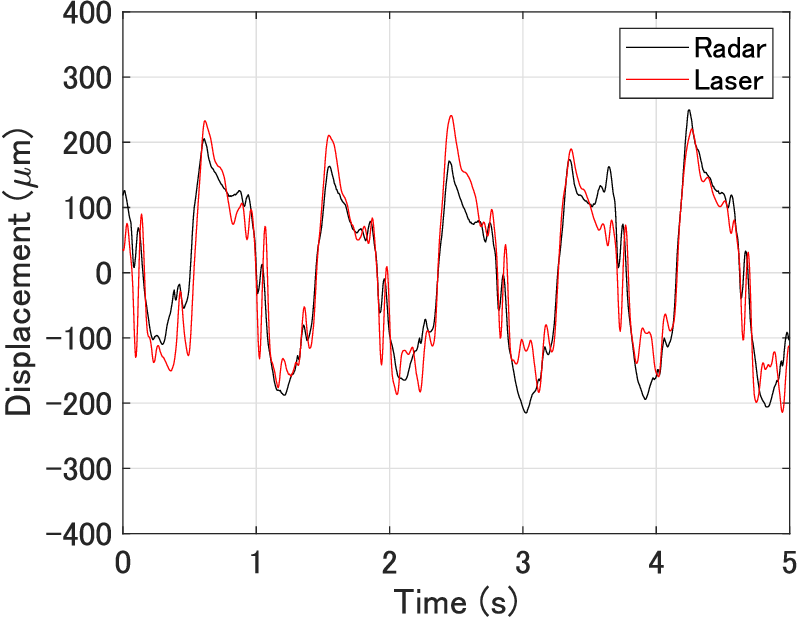}
    \end{minipage}
    \begin{minipage}{0.49\columnwidth}
      \includegraphics[width=\columnwidth]{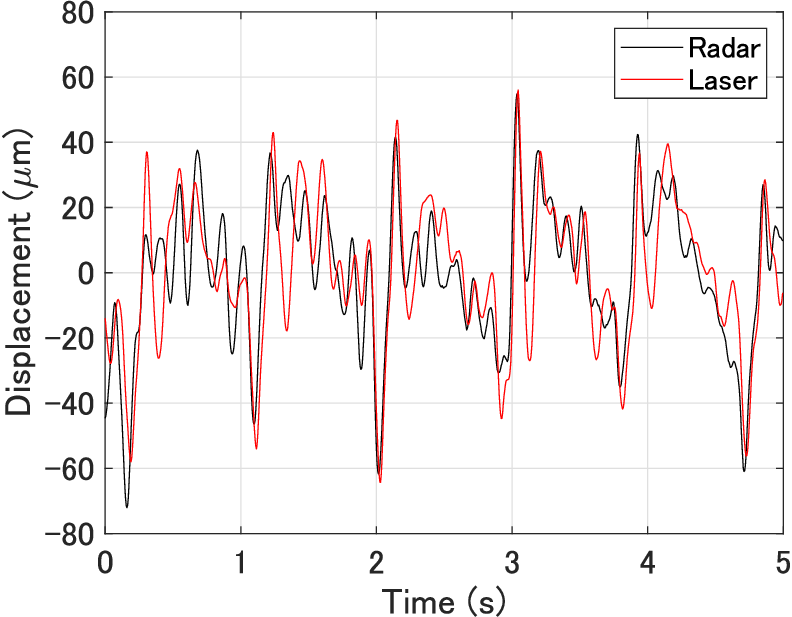}
    \end{minipage}
    \caption{Estimated and reference waveforms of the chest (left) and thigh (right) measured with a radar (black line) and a laser sensor (red line) for participant A in a supine position when the radar position is optimized.}
  \label{fig:5}
\end{figure}

\begin{figure}
    \centering
    \begin{minipage}{0.49\columnwidth}
      \includegraphics[width=\columnwidth]{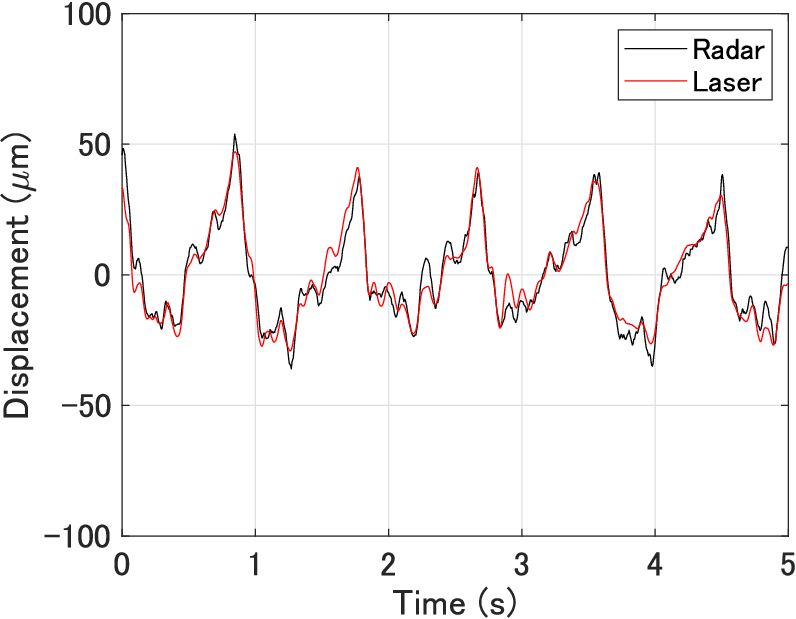}
    \end{minipage}
    \begin{minipage}{0.49\columnwidth}
      \includegraphics[width=\columnwidth]{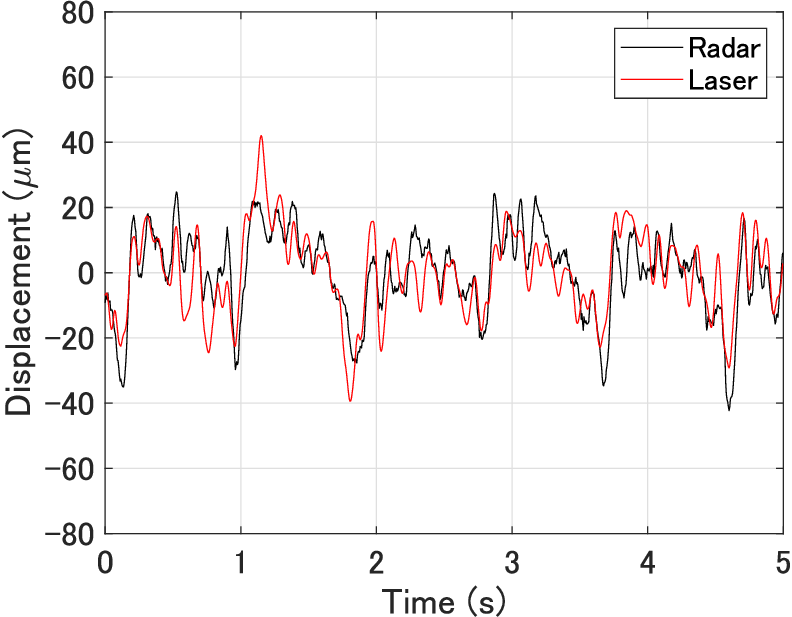}
    \end{minipage}
    \caption{Estimated and reference waveforms of the back (left) and thigh (right) measured with a radar (black line) and a laser sensor (red line) for participant A in a prone position when the radar position is optimized.}
  \label{fig:7}
\end{figure}

\begin{figure}
    \centering
    \begin{minipage}{0.49\columnwidth}
      \includegraphics[width=\columnwidth]{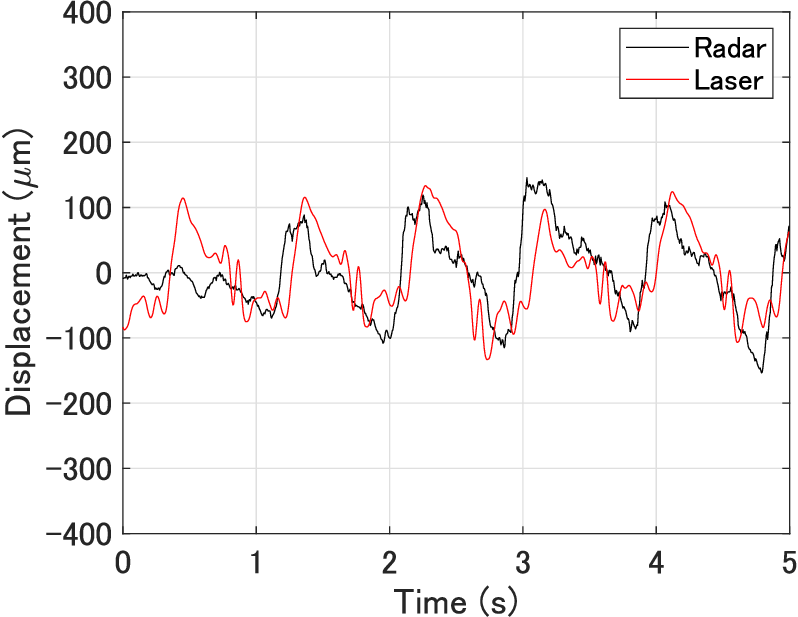}
    \end{minipage}
    \begin{minipage}{0.49\columnwidth}
      \includegraphics[width=\columnwidth]{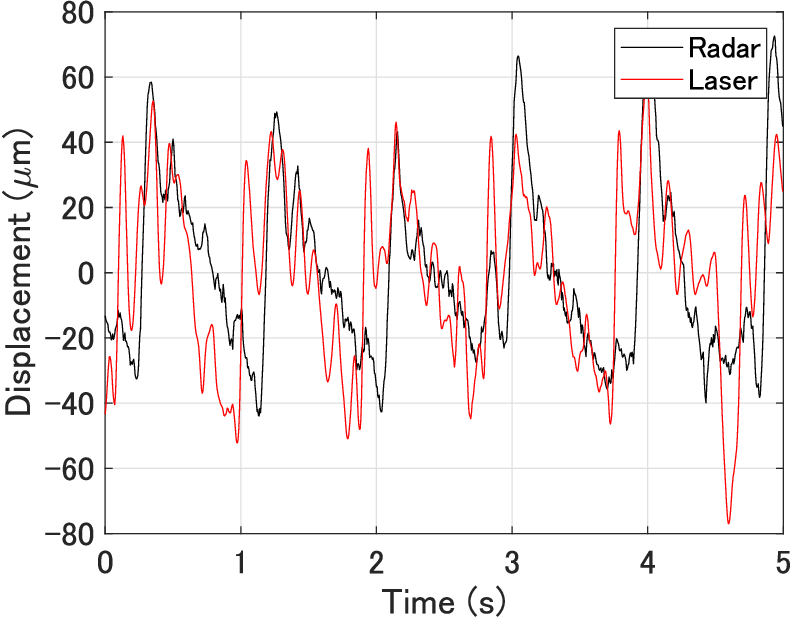}
    \end{minipage}
    \caption{Estimated and reference waveforms of the chest (left) and thigh (right) measured with a radar (black line) and a laser sensor (red line) for participant A in a supine position when the radar position is not optimized.}
  \label{fig:6}
\end{figure}

\begin{figure}
    \centering
    \begin{minipage}{0.49\columnwidth}
      \includegraphics[width=\columnwidth]{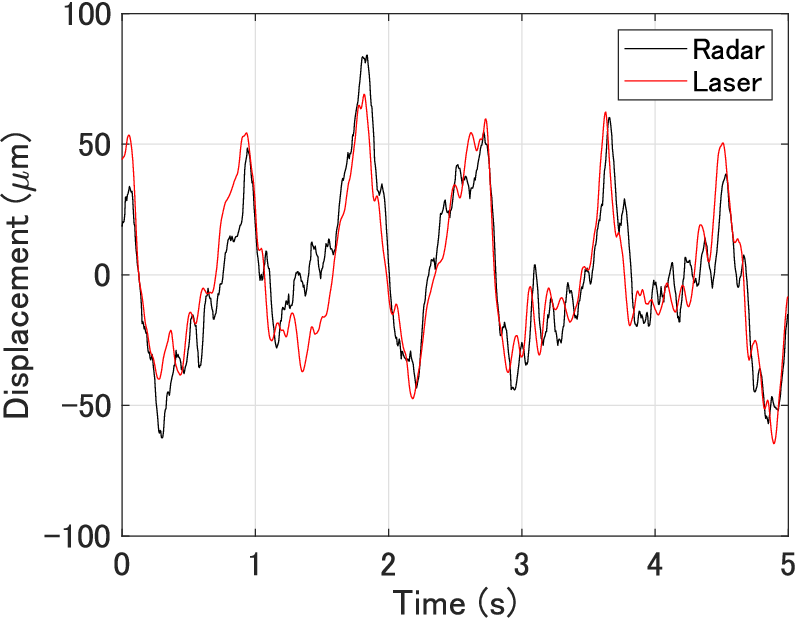}
    \end{minipage}
    \begin{minipage}{0.49\columnwidth}
      \includegraphics[width=\columnwidth]{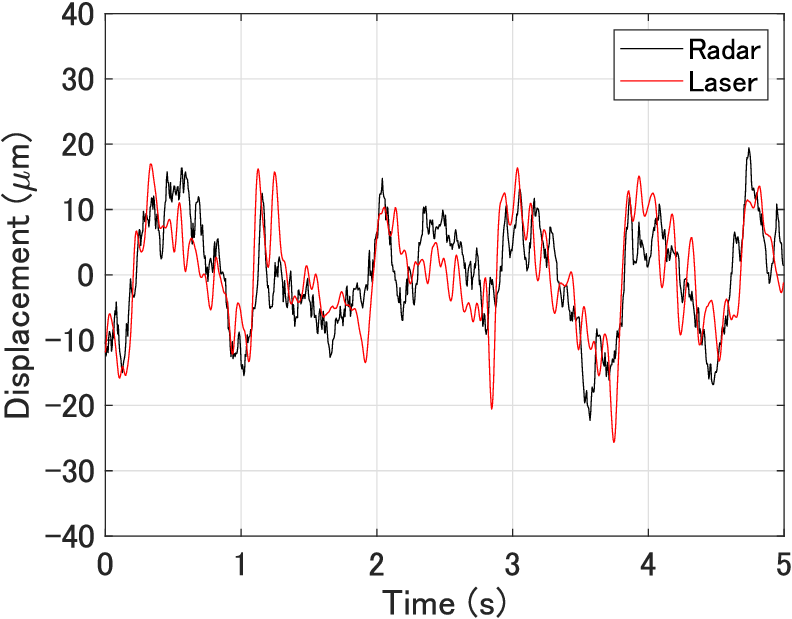}
    \end{minipage}
    \caption{Estimated and reference waveforms of the back (left) and thigh (right) measured with a radar (black line) and a laser sensor (red line) for participant A in a prone position when the radar position is not optimized.}
  \label{fig:8}
\end{figure}

Figs. \ref{fig:5} and \ref{fig:7} show the estimated displacements when the radar position was adjusted using the proposed method for participant A in supine and prone positions, respectively. Here, the black and red lines represent the estimated and reference displacements, respectively. Figs. \ref{fig:6} and \ref{fig:8} show the estimated displacements when the radar position was shifted by $0.3$ m from the case in Figs. \ref{fig:5} and \ref{fig:7}.

First, Table \ref{tab:5} summarizes $\rho_n$ and $\varepsilon_n$ for six measurements of participant A in a supine position, where $i=1$ and $2$ correspond to the chest and front of the thigh, respectively. The average correlation coefficients were $0.66$ and $0.85$ for the conventional and proposed methods, respectively. The average RMS errors were $26.3$ \textmu m and $30.5$ \textmu m for the conventional and proposed methods, respectively. Although the RMS error of the proposed method was 16\% larger than that of the conventional method, the correlation coefficient of the proposed method was improved by 19\% compared with that of the conventional method. 

Next, Table \ref{tab:6} summarizes results for participant B in a supine position. The average correlation coefficients were $0.56$ and $0.87$ for the conventional and proposed methods, respectively. The average RMS errors were $52.3$ \textmu m and $18.6$ \textmu m for the conventional and proposed methods, respectively. The proposed method improved the correlation coefficient and RMS error by factors of 1.6 and 2.8, respectively.

Then, Table \ref{tab:7} summarizes results for participant A in a prone position. The average correlation coefficients were 0.78 and 0.88 for the conventional and proposed methods, respectively. The average RMS errors were $14.2$ \textmu m and $6.2$ \textmu m for the conventional and proposed methods, respectively. The proposed method improved the correlation coefficient and RMS error by factors of 1.1 and 2.3, respectively.

Finally, Table \ref{tab:8} summarizes results for participant B in a prone position. The average correlation coefficients were $0.79$ and $0.75$ for the conventional and proposed methods, respectively. The average RMS errors were $15.3$ \textmu m and $19.5$ \textmu m for the conventional and proposed methods, respectively. In this case, the RMS error of the proposed method was worse than that of the conventional method by a factor of $1.3$, whereas the correlation coefficient was improved by a factor of $1.1$ using the proposed method. In this case, the estimation accuracy was degraded with the proposed method, partly owing to the significantly low accuracy in one of the six measurement datasets. If we exclude the last measurement, the average correlation coefficient and RMS error for the proposed method were $0.93$ and $7.8$~\textmu m, respectively.

\begin{table}[tb]
    \centering
    \caption{Performance evaluation of the proposed method for participant A in a supine position}
    \label{tab:5}
    \begin{tabular}{|c|cccc|} \hline
        Method & $\rho_1$ & $\rho_2$ & $\varepsilon_1$ (\textmu m) & $\varepsilon_2$ (\textmu m) \\ \hline
              & 0.58 & 0.57 & 56.0 & 24.5 \\
        Conventional & 0.82 & 0.69 & 16.9 & 20.1 \\
              & 0.51 & 0.80 & 21.4 & 18.8 \\ \hline
                 & 0.93 & 0.79 & 46.8 & 14.0 \\
        Proposed & 0.82 & 0.71 & 77.3 & 16.3 \\
                 & 0.91 & 0.93 & 21.9 & 6.9 \\ \hline
    \end{tabular}
\end{table}

\begin{table}[tb]
    \centering
    \caption{Performance evaluation of the proposed method for participant B in a supine position.}
    \label{tab:6}
    \begin{tabular}{|c|cccc|} \hline
        Method & $r_1$ & $r_2$ & $e_1$ (\textmu m) & $e_2$ (\textmu m) \\ \hline
              & 0.74 & 0.42 & 51.3 & 23.7 \\
        Conventional & 0.63 & 0.44 & 76.1 & 83.4 \\
              & 0.61 & 0.54 & 54.9 & 24.2 \\ \hline
                 & 0.87 & 0.85 & 38.9 & 9.2 \\
        Proposed & 0.97 & 0.80 & 16.1 & 13.9 \\
                 & 0.91 & 0.82 & 25.5 & 8.4 \\ \hline
    \end{tabular}
\end{table}

\begin{table}[tb]
    \centering
    \caption{Performance evaluation of the proposed method for participant A in a prone position.}
    \label{tab:7}
    \begin{tabular}{|c|cccc|} \hline
        Method & $\rho_1$ & $\rho_2$ & $\varepsilon_1$ (\textmu m) & $\varepsilon_2$ (\textmu m) \\ \hline
              & 0.88 & 0.70 & 14.2 & 6.2 \\
        Conventional & 0.87 & 0.70 & 22.6 & 14.1 \\
              & 0.88 & 0.63 & 13.8 & 14.7 \\ \hline
                 & 0.96 & 0.76 & 4.9 & 8.9 \\
        Proposed & 0.97 & 0.80 & 4.1 & 8.8  \\
                 & 0.93 & 0.86 & 4.3 & 6.1 \\ \hline
    \end{tabular}
\end{table}

\begin{table}[tb]
    \centering
    \caption{Performance evaluation of the proposed method for participant B in a prone position}
    \label{tab:8}
    \begin{tabular}{|c|cccc|} \hline
        Method   & $\rho_1$ & $\rho_2$ & $\varepsilon_1$ (\textmu m) & $\varepsilon_2$ (\textmu m) \\ \hline
              & 0.94 & 0.52 & 11.9 & 10.0 \\
        Conventional & 0.88 & 0.80 & 30.1 & 12.9 \\
              & 0.93 & 0.67 & 8.8 & 17.8 \\ \hline
                 & 0.97 & 0.91 & 7.0 & 6.1 \\
        Proposed & 0.96 & 0.87 & 7.2 & 10.7 \\
                 & 0.44 & 0.36 & 69.1 & 17.2 \\ \hline
    \end{tabular}
\end{table}

\section{Conclusion}
In this study, we proposed a radar sensing system and method for measuring heart pulse waves at multiple sites on the human body using a depth camera and electromagnetic wave scattering analysis. To evaluate the performance of the proposed method, we conducted measurement involving two participants in supine and prone positions using both the proposed and conventional methods. The performances of the methods were evaluated in terms of the correlation coefficient and RMS error in estimating displacement waveforms using reference data taken with laser displacement sensors. Compared with the conventional method, the proposed method improved the estimation accuracy of the body displacement at two positions by 14\% and 19\% on average with respect to the correlation coefficient and RMS error, respectively. These results indicate that the proposed method can achieve accurate measurement of pulse waves at multiple sites. It will be important to study and develop a method for automatic optimization of the radar position using the proposed approach.

\section*{Ethics Declarations}
The experimental protocol involving human participants was approved by the Ethics Committee of the Graduate School of Engineering, Kyoto University (permit no.~201916). Informed consent was obtained from all human participants in the study.

\newpage

\begin{IEEEbiography}
    [{\includegraphics[width=1in,height=1.25in,clip,keepaspectratio]{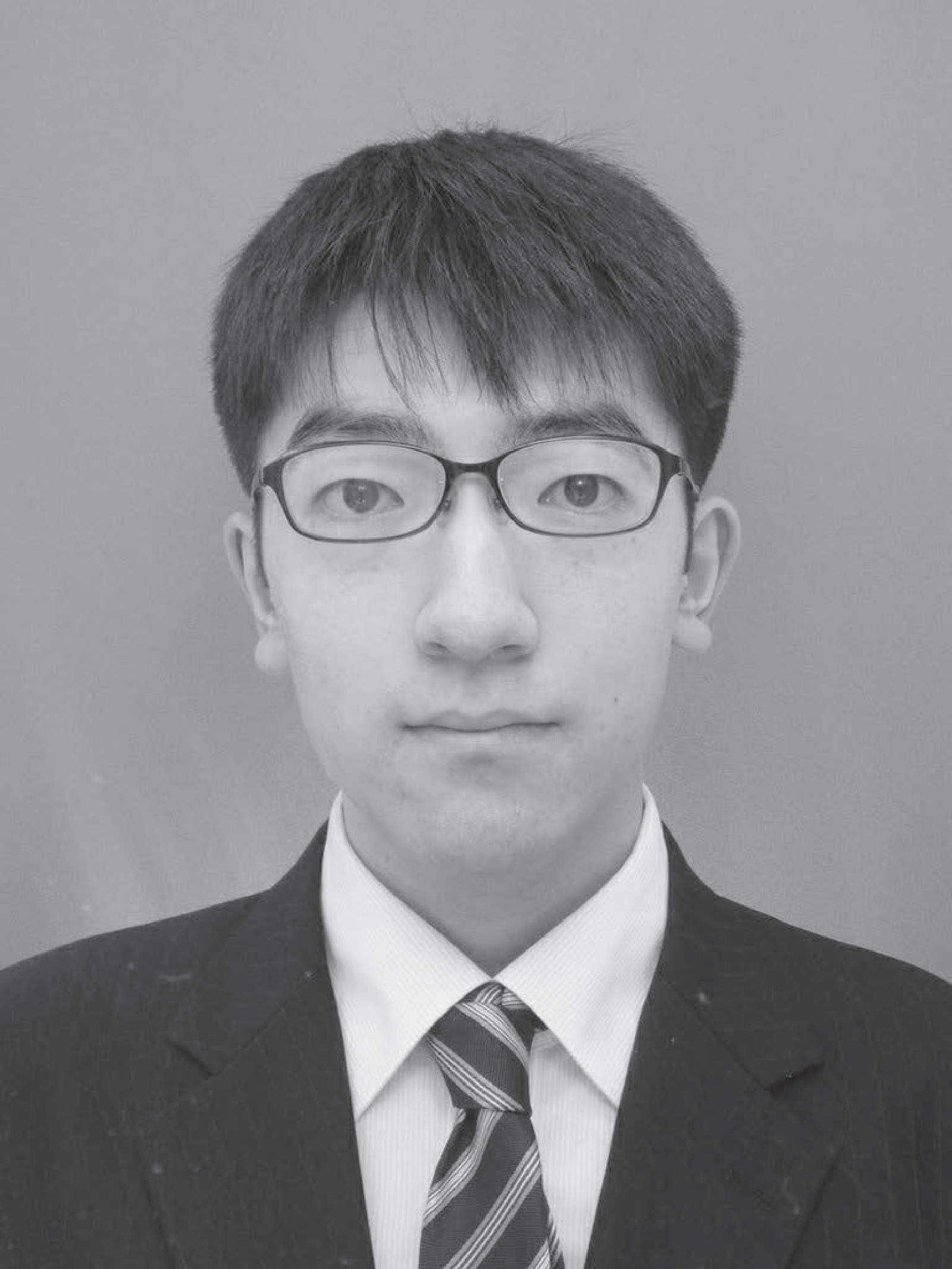}}]{Takehito Koshisaka} 
    received the B.E. degree in electrical and electronic engineering from Kyoto University, Kyoto, Japan, in 2021 and the M.E. degree in electrical engineering from the Graduate School of Engineering, Kyoto University, in 2023.
\end{IEEEbiography}
\newpage
\begin{IEEEbiography}
    [{\includegraphics[width=1in,height=1.25in,clip,keepaspectratio]{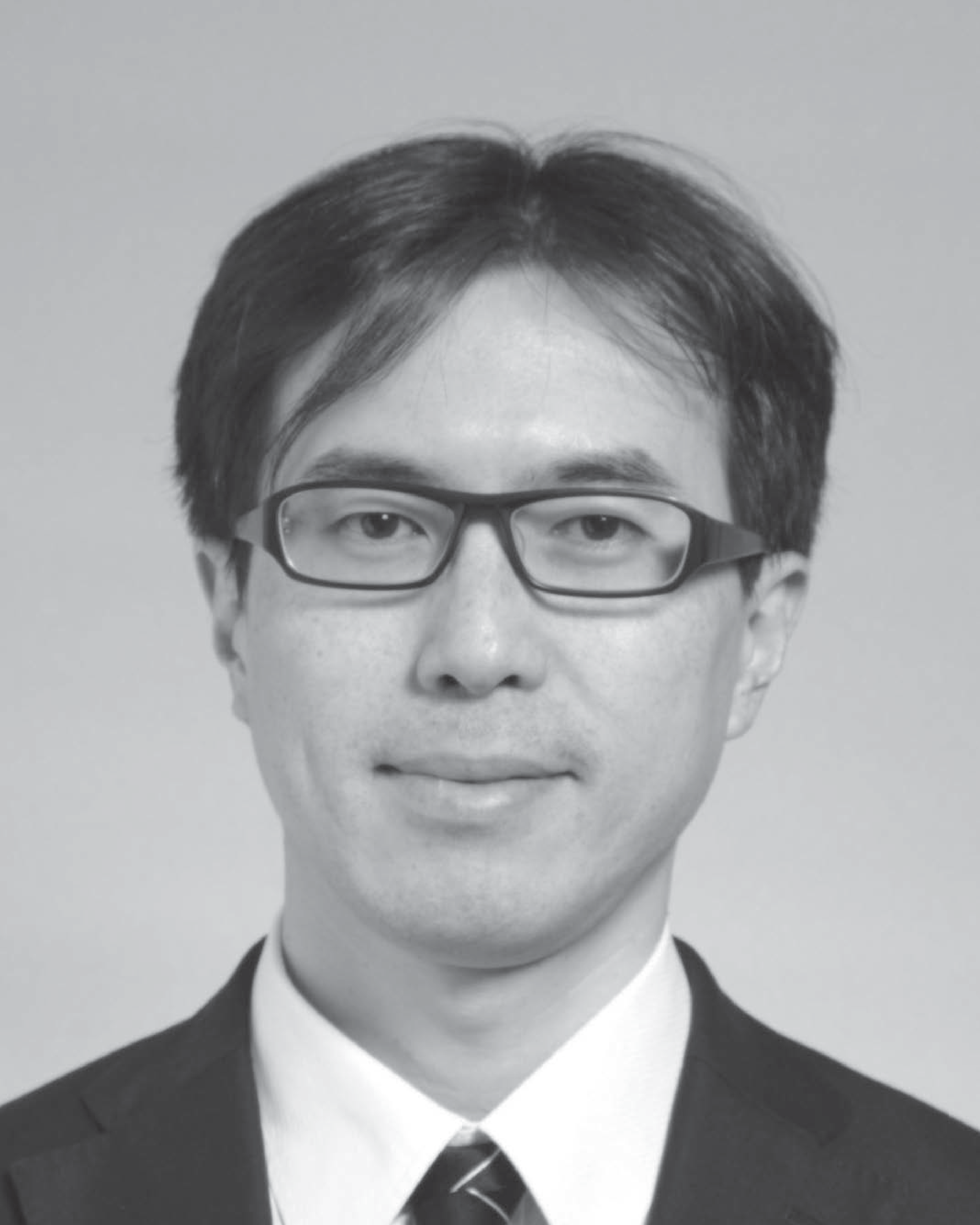}}]{Takuya Sakamoto} (Senior Member, IEEE)
received a B.E. degree in electrical and electronic engineering from Kyoto University, Kyoto, Japan, in 2000 and M.I. and Ph.D. degrees in communications and computer engineering from the Graduate School of Informatics, Kyoto University, in 2002 and 2005, respectively.
  
From 2006 through 2015, he was an Assistant Professor at the Graduate School of Informatics, Kyoto University. From 2011 through 2013, he was also a Visiting Researcher at Delft University of Technology, Delft, the Netherlands. From 2015 until 2019, he was an Associate Professor at the Graduate School of Engineering, University of Hyogo, Himeji, Japan. In 2017, he was also a Visiting Scholar at University of Hawaii at Manoa, Honolulu, HI, USA. From 2019 until 2022, he was an Associate Professor at the Graduate School of Engineering, Kyoto University. From 2018 through 2022, he was also a PRESTO researcher of the Japan Science and Technology Agency, Japan. Since 2022, he has been a Professor at the Graduate School of Engineering, Kyoto University. His current research interests lie in wireless human sensing, radar signal processing, and radar measurement of physiological signals.

Prof. Sakamoto was a recipient of the Best Paper Award from the International Symposium on Antennas and Propagation (ISAP) in 2004; the Young Researcher's Award from the Information and Communication Engineers of Japan (IEICE) in 2007; the Best Presentation Award from the Institute of Electrical Engineers of Japan in 2007; the Best Paper Award from the ISAP in 2012; the Achievement Award from the IEICE Communications Society in 2015, 2018, and 2023; the Achievement Award from the IEICE Electronics Society in 2019; the Masao Horiba Award in 2016; the Best Presentation Award from the IEICE Technical Committee on Electronics Simulation Technology in 2022; the Telecom System Technology Award from the Telecommunications Advancement Foundation in 2022; and the Best Paper Award from the IEICE Communication Society in 2007 and 2023. 
\end{IEEEbiography}

\end{document}